# Mean-squared Energy Difference for Exploring Potential Energy Landscapes of Supercooled Liquids

D. M. Zhang[a], D. Y. Sun[b§] and X. G. Gong[a§]

[a]Key Laboratory for Computational Physical Sciences (MOE), Institute of Computational Physics, Fudan University, Shanghai 200433, China

[b]School of Physics and Electronic Science, East China Normal University, 200241 Shanghai, China

## Abstract

By extending the concept of diffusion to the potential energy landscapes (PELs), we introduce the mean-squared energy difference (MSED) as a novel quantity to investigate the intrinsic properties of glass. MSED can provide a clear description of the "energy relaxation" process on a PEL. Through MSED analysis, we can obtain characteristic timescale similar to those from structure analysis, namely $\tau_\alpha^*$. We establish a connection between MSED and the properties of PELs, providing a concise and quantitative description of the PEL. We find that the roughness of the accessible PEL has changed significantly after the glass transition. And we also find that one of the PEL parameters is closely related to the Adam-Gibbs configurational entropy. The present research, which directly links the PEL to the relaxation process, provides avenues for further research of the glass.

§Corresponding Authors: xggong@fudan.edu.cn; dysun@phy.ecnu.edu.cn

# I. Introduction

Glasses exhibit complex dynamic processes, setting them apart fundamentally from liquids and solids. Unveiling the intrinsic physical mechanisms behind these relaxation processes is essential for demystifying the mysteries of glasses [1-9]. During the formation of glass, the relaxation time can span up to 12 orders of magnitude [10-13]. Such a large range not only poses a challenge to experimental measurement, but also brings great difficulty to theoretical research. For this reason, despite extensive studies on the relaxation process, its essence remains debated and continues to be a focal and challenging topic in glass research.

The two most important and widely studied relaxation processes in glasses are the $\alpha$-relaxation and $\beta$-relaxation [14-18], which were discovered in the energy dissipation spectrum of glasses. Experimentally, the dynamical relaxation behavior of supercooled liquids and glasses is typically measured using dielectric spectroscopy or dynamic mechanical analysis depending on whether the sample has significant dielectric strength. These methods involve applying alternating electric fields (strain) at different frequencies and measuring the dielectric spectrum (dynamic modulus) of the sample. If a corresponding relaxation process exists at a given frequency, the dissipation intensity at that frequency increases. It has been observed that as temperature decreases, two broad peaks appear in the energy dissipation spectrum of supercooled liquids, corresponding to two relaxation modes. The slow and fast modes refer to $\alpha$-relaxation and $\beta$-relaxation, respectively [14,19-23]. In some materials, however, the spectrum may exhibit a main peak and an excess wing instead of two peaks [24-26], and the microscopic origin of the excess wing is still under debate [19]. Nevertheless, from a fundamental perspective, all relaxation processes should be determined by the morphology of the potential energy landscapes (PEL), i.e., the system navigating on a high-dimension PEL. A widely adopted PEL picture is that $\beta$-relaxation corresponds to transitions between local minima (i.e., basins), while $\alpha$-relaxation refers to transitions between meta-basins (MBs) [27-33]. As pointed out previously [34,35], such picture should carefully consider the spatial dynamic heterogeneity, especially for larger systems.

Despite substantial experimental and theoretical research on glass relaxations, many important issues remain unresolved. Especially, relaxation times from different theoretical methods can vary significantly, especially at high temperatures (e.g., Ref [36]). In typical theoretical simulations, $\alpha$-relaxation time ($\tau_\alpha$) and $\beta$-relaxation time ($\tau_\beta$) are derived from specific structural correlation functions, with energy information entirely discarded. While in experiments, it is difficult to directly observe the micro-structure of glasses, so energy dissipation is often measured instead. However, the relaxation times derived from structural relaxation in computer simulations and energy dissipation in experiment lack direct connections, and it is unclear if they correspond to the same relaxation process. Especially, since $\alpha$-relaxation is considered as the process that the system "walking" on a PEL, can one obtain valuable information about the PEL from $\alpha$-relaxation? Besides, the methods based on structural correlation functions often involve significant computational costs. Is there a simple approach to calculate relaxation times? These questions motivate us to explore a faster, energy-based method for studying relaxation process.

We note that changes in potential energy of a system over time reflect the relaxation process on the PEL. Undoubtedly, all processes of the system on the PEL should be reflected in the time series of potential energy. However, the valuable information is obscured by random thermal fluctuations [37], which is much difficult to extract from a time series of potential energy. Simple thermodynamic averaging may remove thermal fluctuations, but it can also erase relaxation-related information. Conversely, simply considering the time variation of potential energy lacks statistical reliability. In this work, we introduce a method that provides sufficient time averaging without losing time information. Specifically, we analogize the "motion" of a system on the PEL to atomic motion in real space, treating the energy changes as a form of "diffusion" behavior. Through systematic analysis, we clearly describe the system's "walking" process on the PEL and obtain the $\alpha$-relaxation time. Additionally, we can derive some quantitative information describing the morphology of PELs from the relaxation processes, which could be useful for calculating configurational entropy. This paper not only establishes a simple algorithm for studying relaxation processes, but also provides a new perspective on the relationship between relaxations and the PEL of glasses.

## II.  Methodological Development

Einstein has derived the relationship between the diffusion coefficient and the mean-squared displacement (MSD) starting from the Brownian motion of particles [38]. Here, we follow his lead and extend this analogy to energy spaces, deriving a corresponding energy "diffusion equation".

First, denote $\Delta$ as the energy change after a short time interval $\tau$, and the probability distribution of $\Delta$ as $\varphi(\Delta)$. Next, we denote the probability distribution function of energy $E$ at time $t$ as $f(E,t)$. After the short time interval $\tau$, the evolved probability distribution function $f(E, t + \tau)$ can be obtained from the basic principles of probability statistics:

$$f(E, t + \tau) = \int_{-\infty}^{+\infty} f(E + \Delta, t)\varphi(\Delta)d\Delta. \tag{1}$$

To derive a differential equation, we apply a Taylor expansion to the left-hand side of Eq. (1):

$$f(E, t + \tau) = f(E, t) + \tau \frac{\partial f(E, t)}{\partial t} + \cdots. \tag{2}$$

And in the right-hand side of Eq. (1):

$$f(E + \Delta, t) = f(E, t) + \Delta \frac{\partial f(E, t)}{\partial E} + \frac{\Delta^2}{2} \frac{\partial^2 f(E, t)}{\partial E^2} + \cdots. \tag{3}$$

Neglecting higher-order terms and substituting Eqs. (2) and (3) into Eq. (1), we obtain:

$$f(E, \delta t) + \tau \frac{\partial f(E, t)}{\partial t} = f(E, t) \int_{-\infty}^{+\infty} \varphi(\Delta)d\Delta + \frac{\partial f(E, t)}{\partial E} \int_{-\infty}^{+\infty} \Delta \varphi(\Delta)d\Delta$$
$$+ \frac{\partial^2 f(E, t)}{\partial E^2} \int_{-\infty}^{+\infty} \frac{\Delta^2}{2} \varphi(\Delta)d\Delta. \tag{4}$$

Clearly, since

$$\int_{-\infty}^{+\infty} \varphi(\Delta)d\Delta = 1, \tag{5}$$

we finally get:

$$\frac{\partial f}{\partial t} = \frac{\langle \Delta \rangle}{\tau} \frac{\partial f}{\partial E} + \frac{\langle \Delta^2/2 \rangle}{\tau} \frac{\partial^2 f}{\partial E^2}, \tag{6}$$

where $\langle \Delta \rangle$ represents the change in energy over $\tau$. If the system is in thermal equilibrium, we have $\langle E \rangle \equiv const.$, which is equivalent to:

$$\int_{-\infty}^{+\infty} \Delta \varphi(\Delta)d\Delta = \langle \Delta \rangle = 0. \tag{7}$$

Let $D_E = \frac{1}{\tau} \int_{-\infty}^{+\infty} \frac{\Delta^2}{2} \varphi(\Delta)d\Delta$, and we obtain the energy diffusion equation in equilibrium systems:

$$\frac{\partial f}{\partial t} = D_E \frac{\partial^2 f}{\partial E^2}, \tag{8}$$

where $D_E$ is called the energy diffusion coefficient. $\langle \Delta^2 \rangle$ is the mean-squared energy difference (MSED), corresponding to MSD in real space. Since the energy of the system is a one-dimension variable, $\langle \Delta^2 \rangle$ will eventually reach a certain value in equilibrium. From the definition of $\Delta$, we know that this value corresponds to the energy fluctuation:

$$\lim_{t \to \infty} \langle \Delta^2 \rangle = \langle E^2 \rangle - \langle E \rangle^2 = N k_B T^2 C_V. \tag{9}$$

Thus, beyond a sufficiently long period of time, $D_E$ will approach zero in the form of $t^{-1}$, indicating that the energy distribution stabilizes and no longer changes, meaning the system has "seen" the complete PEL it can explore. With the above mathematical derivation and the results shown in this paper, we will illustrate the feasibility of considering energy evolution as a diffusive behavior, which may provide some inspiration in glass researches. In this paper, we will focus on the behavior of MSED, compared to $D_E$, it provides a more intuitive picture of the system "walking" on the PEL.

## III.  Computational Details

In this study, we investigate supercooled liquids of the $Cu_{50}Zr_{50}$ and Kob-Anderson binary Lennard-Jones (KA-BLJ) systems. All molecular dynamics (MD) simulations were performed using the LAMMPS code [39]. The $Cu_{50}Zr_{50}$ system exhibits a strong glass-forming ability and remains stable throughout the supercooled temperature range. We simulated a bulk system containing 2048 atoms using the optimized EAM potential for supercooled liquids from Mendelev et al. [40]. Initially, 20 different configurations were prepared at 2000 K, followed by cooling at a rate of $10^{10}$ K/s to 300 K. The entire cooling process was carried out using the NPT ensemble at zero external pressure. In our simulations, the glass transition temperature $T_g$ for $Cu_{50}Zr_{50}$ is approximately 760 K, which is in agreement with previous studies [41,42]. At the temperatures of interest, the system underwent 1 ns NVT relaxation before changing to the NVE ensemble for sampling, with a sampling duration of up to 400 ns for each temperature.

In contrast, the KA-BLJ system undergoes a glass transition without the need for rapid cooling. We simulated a KA-BLJ system containing 1000 atoms (with a ratio of $N_A:N_B = 80:20$) using the NVT ensemble, cooling from $T = 1.0$ to $0.1$ at a rate of $6 \times 10^{-7} \tau^{-1}$. The glass transition temperature for KA-BLJ is $T_g = 0.39$. At each temperature of interest, long NVE ensemble simulations were conducted for up to $10^7 \tau$. The simulations were carried out using reduced units, with parameters consistent with [43]. For comparison, the reduced units are converted to real units for the $Ni_{80}P_{20}$ system in the following content, and the conversion factors are: energy $\epsilon = 0.0804$ eV, temperature $T = 932.82$ K, and time unit $\tau = 6.2721 \times 10^{-13}$ s.

We calculated the MSD for both systems at temperatures of interest. MSD is simply defined as $\langle \Delta r(t)^2 \rangle = \langle (r(t) - r(0))^2 \rangle$, in which $r(0)$ and $r(t)$ are the atom coordinates at time 0 and t. As for energy, to ensure that all results are independent of the number of atoms, we define $\Delta$ as the difference in average atomic potential energy, i.e., $\Delta(t) = \frac{1}{N}[E(t_0) - E(t_0 + t)]$, which does not alter the behavior of $\langle \Delta(t)^2 \rangle$. MSED was then calculated by

$$\langle \Delta(t)^2 \rangle = N^{-2} \langle (E(t) - E(0))^2 \rangle. \tag{10}$$

To obtain the α-relaxation time $\tau_\alpha$, we computed the self-intermediate scattering function (SISF) $F_s(q = q_{max}, t) = \langle \exp\{-i q \cdot (r(0) - r(t))\} \rangle$, in which $q_{max}$ is the wave vector where the static structure factor $S(q)$ reaches its first maximum [44,45]. In previous researches, $\tau_\alpha$ is usually defined as the duration required for SISF to decay from 1 to $e^{-1}$.

Our study primarily focused on the system's behavior within the supercooled liquid temperature range. The entire simulation process was carefully monitored to ensure no crystallization occurred during sampling.

## IV. Results and Discussion

Fig. 1 shows $\varphi(\Delta)$ for different temperatures and durations in $Cu_{50}Zr_{50}$ system. One can see that $\varphi(\Delta)$ has almost the same form as the distribution of atomic displacements, namely a Gaussian distribution. At higher temperature (e.g., 900 K in Fig. 1(a)), since the system explores the PEL relatively fast, $\varphi(\Delta)$ shows an evident change from $t = 1$ ps (red curve) to 10 ps (blue curve), and remains almost unchanged after 100 ps (green and black curves). However, at $T = 800$ K (Fig. 1(b)), $\varphi(\Delta)$ only changes slightly from $t = 1$ ps to 10 ps, and keeps evolving in longer time. The comparison indicates the feasibility of considering energy evolution as a diffusive behavior.

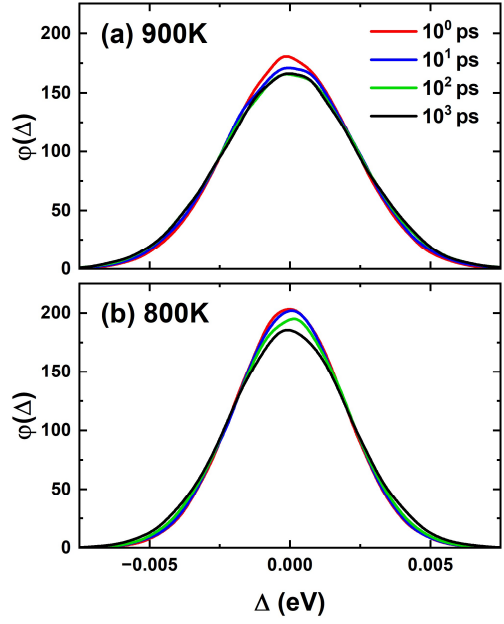

**Figure 1.** The probability distribution function $\varphi(\Delta)$ in $Cu_{50}Zr_{50}$ system over different time intervals at (a) 900 K and (b) 800 K. $\varphi(\Delta)$ is nonzero only in the vicinity of $\Delta = 0$. As $t$ increases, the system explores larger regions of the PEL, and the width of the distribution expands gradually. At $T = 900$ K, $\varphi(\Delta)$ changes evidently from $t = 1$ ps to $10$ ps, and remains almost the same after $t = 100$ ps. While at $T = 800$ K, $\varphi(\Delta)$ changes very slightly from $t = 1$ ps to $10$ ps, and keeps evolving from $t = 100$ ps to 1 ns.

Fig. 2 shows both MSD and MSED in $Cu_{50}Zr_{50}$ system. Based on previous analyses about MSD and the SISF, atomic motion in supercooled liquids can generally be divided into three stages: free motion, confined motion within a local "cage", and diffusive motion [6,46-49]. The MSED perfectly reflects these stages of structural evolution. It can be seen that MSED exhibits a peak in a very short time (around 0.05 ps), followed by a decay to the first plateau around 0.2 ps. Since we are calculating the square of energy changes, this peak actually corresponds to oscillations in the total energy of the system with a period of 0.1ps (the regular oscillations seen within 1.0 ps in the figure). This time period is almost independent of temperature and corresponds to the characteristic time of atomic vibrations. For $t < 0.1$ ps, the change in MSED reflects the vibrations of systems near the potential energy minima. Thus, for $t < 0.1$ ps, both MSD and MSED are proportional to $t^2$ (note that MSD in Fig. 2(a) is plotted on a double logarithmic scale). In the following discussion, we will no longer focus on this timescale.

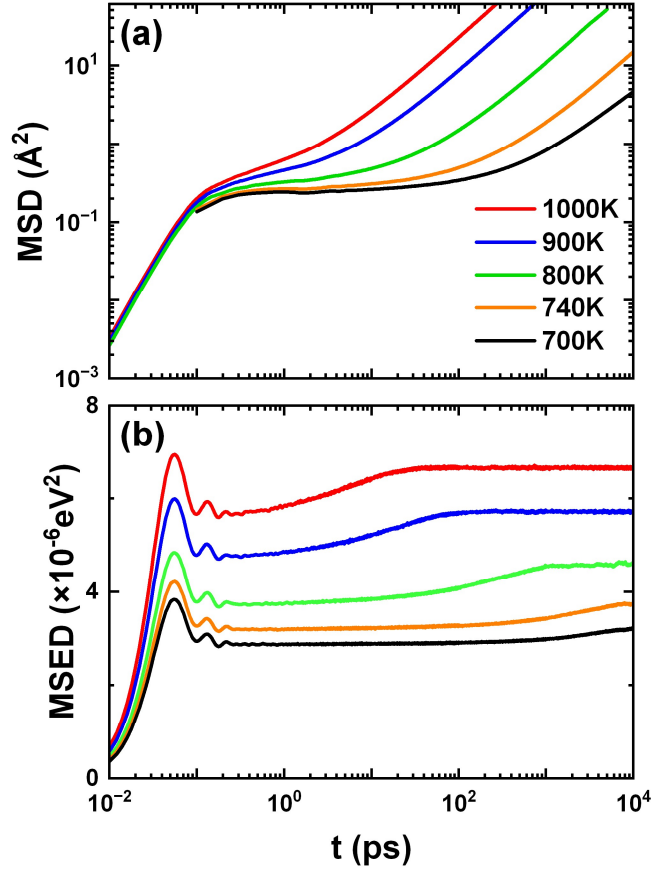

**Figure 2. Mean-squared displacement (a) and Mean-squared energy difference (b) for Cu$_{50}$Zr$_{50}$ at several temperatures. It can be seen that the information about atomic vibration (the regular oscillation within 1 ps in (b)) was eliminated in MSD while kept in MSD. Both functions show plateau in intermediate time, corresponding to the cage breaking regime in real space and intra-meta-basin hopping in potential energy landscapes, respectively. Over time, MSD eventually increases proportional to *t* in diffusion regime, while MSED grows and reaches platform after the system "sees" all the accessible areas on potential energy landscapes.**

After the ballistic regime, the motion of atoms in real space becomes constrained by their neighboring atoms, causing the MSD reaches a local maximum (though not very pronounced), then enters an almost plateau. The time that the MSD reaches the local maximum is defined as $\tau_\beta$ in literatures (e.g. [16]). At low temperatures, atoms remain trapped in a "cage" for longer periods (this may originate from the tendency of atoms moving more cooperatively [6,50,51]), which extends the plateau. From the PEL perspective, this timescale corresponds to the system's motion between different local minima, commonly referred to as the $\beta$ regime. We observe that the MSED does not exhibit obvious oscillations in the $\beta$ regime. The absence of the $\beta$-relaxation signal in MSED may stem from the fact that $\beta$ events are localized, with no significant correlation or cooperation among them [19,52,53]. Since $\beta$-relaxation involves local motion of a small number of atoms and occurs over short timescales, multiple $\beta$-relaxation events can exist simultaneously in the system [54,55]. This randomness causes different $\beta$-relaxation processes to cancel each other out in terms of their impact on the total energy.

On longer timescales, the MSED enters a region where it first increases steadily and then approaches saturation. In the early stage of this process, as the system explores larger regions of the PEL, the MSED continues to grow slowly. Eventually, when the system "sees" the full extent of the accessible PEL, the MSED stabilizes at a certain value, consistent with our earlier analysis. During this time window, particle motion in real space transitions into the diffusion stage, where MSD shifts from the plateau to proportionality with $t$. At this timescale, the slow change in MSED indicates that the involved relaxation times are long. Based on this, we can preliminarily conclude that the system begins transitioning between MBs, which likely corresponds to the $\alpha$-relaxation process. We will verify this conclusion in the following analysis.

Unlike thermal vibration and $\beta$-relaxation, $\alpha$-relaxation corresponds to the transition between MBs, and the associated relaxation time is sufficiently long. In this case, the change in potential energy mainly reflects the average energy of each MB. When the system is confined within one MB, and only the background strength provided by thermal vibrations is considered, the MSED can be approximated as $\Delta^2(t) = \Delta^2(0) = U_0^2$. Assuming the average energy difference between MBs is $U_m$ and the characteristic transition time is $\tau_\alpha^*$, the probability that the system remains in the same energy well after time $t$ can be written as $P(t) = e^{-\frac{t}{\tau_\alpha^*}}$. Once a transition occurs, the energy difference becomes $\Delta^2(t) = \Delta^2(0) + U_m^2$. Clearly, the time evolution of $\langle \Delta^2(t) \rangle$ can be expressed as:

$$\langle \Delta^2(t) \rangle \cong \Delta^2(0) + U_m^2[1 - P(t)] = U_0^2 + U_m^2\left(1 - e^{-\frac{t}{\tau_\alpha^*}}\right). \tag{11}$$

From this equation, we can see that $\alpha$-relaxation does not cause MSED to exhibit the regular oscillatory behavior. There are two main reasons: 1) The relaxation time of $\alpha$-relaxation follows an exponential distribution, indicating significant temporal heterogeneity of dynamic processes [37,56]. This means that $\alpha$-relaxation events are much less correlated in time, and thus, we do not observe oscillations in MSED; 2) The characteristic timescale for movement between MBs is much longer, and the time the system spends in the saddle point is negligible compared to the time it spends near the minima. As a result, information about barrier heights is effectively averaged out over long time and does not manifest in energy fluctuations.

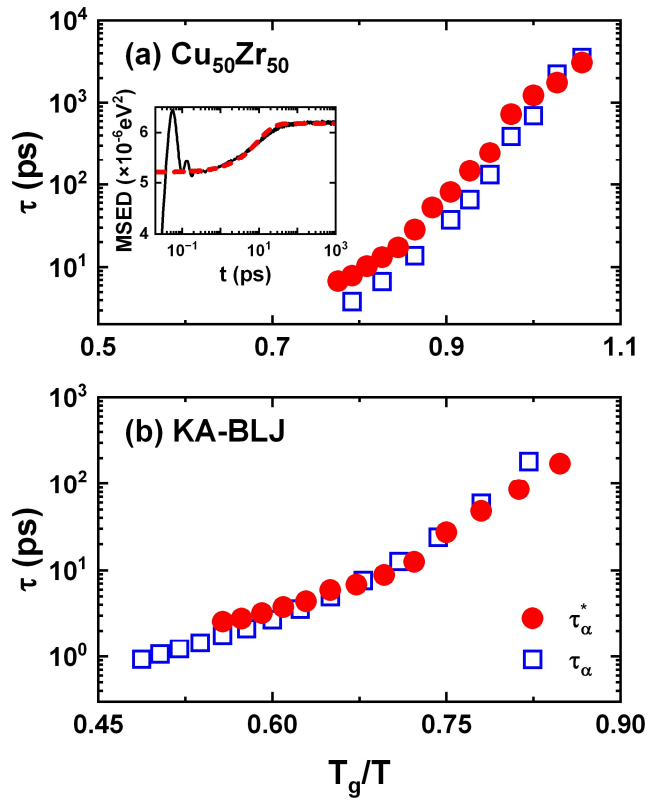

**Figure 3. Relaxation time in (a) $Cu_{50}Zr_{50}$ and (b) KA-BLJ. $\tau_\alpha^*$ (filled circle) is the relaxation time obtained by fitting mean-squared energy difference with Eq. (11), and $\tau_\alpha$ (open square) stands for the structural relaxation time obtained from self-intermediate scattering function, and. Reduced units in KA-BLJ system have been converted to real units (corresponding to $Ni_{80}P_{20}$ system). We can see that $\tau_\alpha$ and $\tau_\alpha^*$ are similar in value and share the same tendency, which suggest the feasibility of obtaining relaxation time from energy correlation function. The inset in (a) shows the mean-squared energy difference in $Cu_{50}Zr_{50}$ at 920 K (black line), and the curve fitted using Eq. (11) (red dashed line). It is clearly shown that Eq. (11) could describe the growing behavior quite well.**

Our fitting results indicate that the long-time behavior in MSED indeed corresponds to the $\alpha$-relaxation process in the system. the relaxation time $\tau_\alpha^*$ is obtained and is plotted in Fig. 3, where (a) corresponds to the $Cu_{50}Zr_{50}$ system and (b) to the KA-BLJ system. For comparison, the $\tau_\alpha$ obtained from the SISF based on structural correlation are also shown in the figure. It is clearly shown that the $\tau_\alpha^*$ obtained from MSED analysis and $\tau_\alpha$ from structure analysis exhibit the same trend and are very close in magnitude. This is in agreement with the work of Baity-Jesi et al., in which they investigated the typical escape time of MBs by studying inherent structures [57]. At higher temperatures, $\tau_\alpha^*$ is slightly larger than $\tau_\alpha$, while at lower temperatures, $\tau_\alpha^*$ becomes slightly smaller than $\tau_\alpha$. Such difference may arise the fact that the MSED is more sensitive to relaxation processes that cause significant energy changes, whereas SISF is more sensitive to processes that cause structural changes. This aligns with the nature of $\alpha$-relaxation, which has a broad distribution and involves complex relaxation processes. Finally, in the $Cu_{50}Zr_{50}$ system, both $\tau_\alpha^*$ and $\tau_\alpha$ exhibit a turning point near the glass transition temperature $T_g$ (around 760K), a phenomenon referred to as the fragile-to-strong crossover [37,58-61].

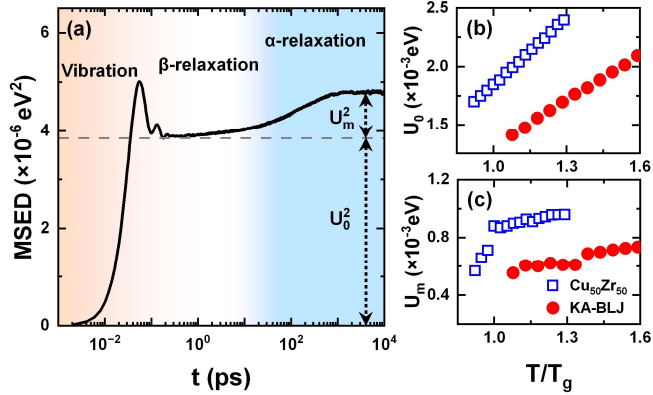

**Figure 4. (a) MSED of $Cu_{50}Zr_{50}$ at 820 K. $U_0$ and $U_m$ are fitting parameters of Eq. (11), and dashed lines and arrows are guides to the eye. It can be seen that MSED can distinguish energy fluctuation contributed by fast and slow processes. According to our analysis, $U_0$ in (b) is the background fluctuation intensity provided by fast processes and is proportional to temperature, while $U_m$ in (c) is the average energy difference between meta-basins. When being cooled to $T_g$, $U_m$ of $Cu_{50}Zr_{50}$ shows a significant turning and decreases faster, which indicates the system moving in a flatter area on potential energy landscape.**

Our theoretical analysis (Eq. (11)) not only provides a model for relaxation time but also gives quantitative information about the PEL. Fig. 4(a) shows a typical MSED curve for $Cu_{50}Zr_{50}$ at 820 K, with dashed lines and arrows serving as guides to the eye. Here, $U_0^2$ represents background fluctuations excluding the $\alpha$-relaxation process, which includes contributions from thermal vibrations and short-term relaxation processes. $U_m$, on the other hand, is the average energy difference between MBs, and its square corresponds to the potential energy fluctuations caused by $\alpha$-relaxation. Based on the definition, we expect $U_0 \propto k_B T$. Figs 4(b) and 4(c) present the fitting results of Eq. (11) for the $Cu_{50}Zr_{50}$ and KA-BLJ systems, respectively. It is clearly shown that $U_0$ is proportional to $T$, which aligns perfectly with our expectations. Since $U_m$ is proportional to the average energy difference between MBs, as the temperature decreases, configurations with higher energies become inaccessible due to reduced kinetic energy. Consequently, we observe a decrease in $U_m$ with decreasing temperature. Interestingly, as the temperature drops, $U_m$ first decreases slowly until it reaches a turning point, after which it declines more rapidly. This transition occurs near $T_g$, the glass transition temperature. This change suggests that below $T_g$, the accessible configuration space shrinks more quickly with cooling. From the perspective of the PEL, this corresponds to a significant reduction in the system's likelihood of accessing MBs with large energy differences after the glass transition, resulting in the system being trapped in a relative flatter region. These results are in agreement with previous studies that the typical energy difference between MBs decreases with temperature, and the relaxation is governed by the rare low-energy paths at lower temperature [57].

Among various theoretical models for $\alpha$-relaxations [62-66], the Adam-Gibbs model stands out as the most influential one and plays a pivotal role in the study of $\alpha$-relaxation [67-69]. The key point in the AG model lies in that $\tau_\alpha$ is governed by the configurational entropy (labeled as $S_{AG}$). The present study may provide a connection between PEL and $S_{AG}$. Considering the excess specific heat $C_{ex} = \frac{\Delta^2}{Nk_B T}$ (where the specific heat only reflects contributions from potential energies), we can obtain the system's excess entropy $S_{ex}$ using $S_{ex} = \int \frac{C_{ex}}{T} dT$. According to Eq (11), for sufficiently long time, the MSED can be expressed as the sum of two components: $\Delta^2 =$

$U_0^2 + U_m^2$. Therefore, the excess specific heat and entropy can be written as: $C_{ex} = \frac{U_0^2}{Nk_BT^2} + \frac{U_m^2}{Nk_BT^2}$, and $S_{ex} = S_1 + S_2$, where:

$$S_1 = \int \frac{U_0^2}{Nk_BT^2} dT, S_2 = \int \frac{U_m^2}{Nk_BT^2} dT. \qquad (12)$$

Since $U_0^2$ mainly reflects contributions from thermal vibrations and relaxations with short time, $S_1$ mainly corresponds to vibrational entropy and the configurational entropy induced by short-time and short-range atomic rearrangements. While $U_m^2$ primarily reflects energy fluctuations due to long-time $\alpha$-relaxation, $S_2$ mainly corresponding to configurational entropy induced by long-time and long-range atomic rearrangements. Thus, $S_2$ could be the configurational entropy discussed in the Adam-Gibbs model, i.e., $S_{AG}$. It needs to be pointed out that, although MSD and MSED are somewhat comparable in their own right, the ideas of computing configurational entropy derived from the two methods are different [70].

## Conclusion

This study introduces the mean-squared energy difference (MSED) as a tool for analyzing the relaxation processes in supercooled liquids. By extending the notion of diffusion to potential energy landscapes (PELs), MSED enables a clear and experimentally relevant depiction of how a system "walking" on PELs, and further gives a characteristic timescale which shares similar values and tendency with structural relaxation time. We demonstrate that MSED offers insights comparable to structural correlation functions but with a more direct connection to energy dissipation. The present study successfully established a relationship between MSED and the PEL, revealing significant changes from a rough PEL to a relatively flat PEL with the glass transition. We show that, MSED presents a framework for studying the dynamical behavior of glassy states.

**Acknowledgements:** D. Y. Sun was supported by the National Natural Science Foundation of China (Grant No. 12274127), and by the National Key Research and Development Program of China (Grant No. 2022YFA1404603). X. G. Gong was supported by NSFC grant (Grant No. 12188101).